# Teaching with Physlets®: Examples from Optics

*Melissa Dancy, Wolfgang Christian and Mario Belloni*
*Davidson College*

**Abstract**: Physlets are scriptable Java applets that can be used for physics instruction. In this article we discuss the pedagogical foundations of Physlet use and provide a sample of Physlet-based exercises that could be used to teach optics.

Research on problem solving[1,2,3,4] indicates that students often approach typical end of chapter problems with little regard to the conceptual underpinnings of the problem. They direct most of their efforts at finding a formula that contains the variables given in the problem statement. A number of strategies[5,6,7,8,9] have been suggested to help students develop a more expert-like approach toward problem solving. These strategies generally guide the student through various steps of solving a problem in an effort to help them learn to conceptualize and define a problem before attempting to plug in numbers. Another approach is to give students alternative problems that discourage novice problem-solving approaches. For example, context-rich problems[10] are phrased in terms of a story, often give students more information than is necessary, and do not explicitly state the variable to be solved for. Another example of alternative problems is estimation problems,[11] which are understated, requiring the student to bring his or her own information to the problem. We use Physlets to create alternative problems that we believe can help students to better develop their problem solving ability and deepen their conceptual understanding.

**The Pedagogy of Physlets**

A Physlet® is a scriptable Java applet developed at Davidson College. There are currently 24 Physlets that have been used by various instructors to script thousands of problems and questions. These problems and questions provide almost complete coverage for introductory physics,[12] as well as topics from upper-level courses such as quantum mechanics.[13] Once a Physlet problem is scripted it is easily delivered over the World Wide Web on an html page. More details about creating Physlet problems, setting up an html page to deliver the problem, and a collection of ready to run problems, can be found in the *Physlets* book[14] or on the Physlets website.[15]

An example of a seemingly straight-forward Physlet-based problem is shown below in Figure 1. In this particular problem students see the applet as shown in the figure and are told, “A point source is located to the left of a mirror. You can drag this point source to any position. Position is given in centimeters. Find the focal length of the mirror.”

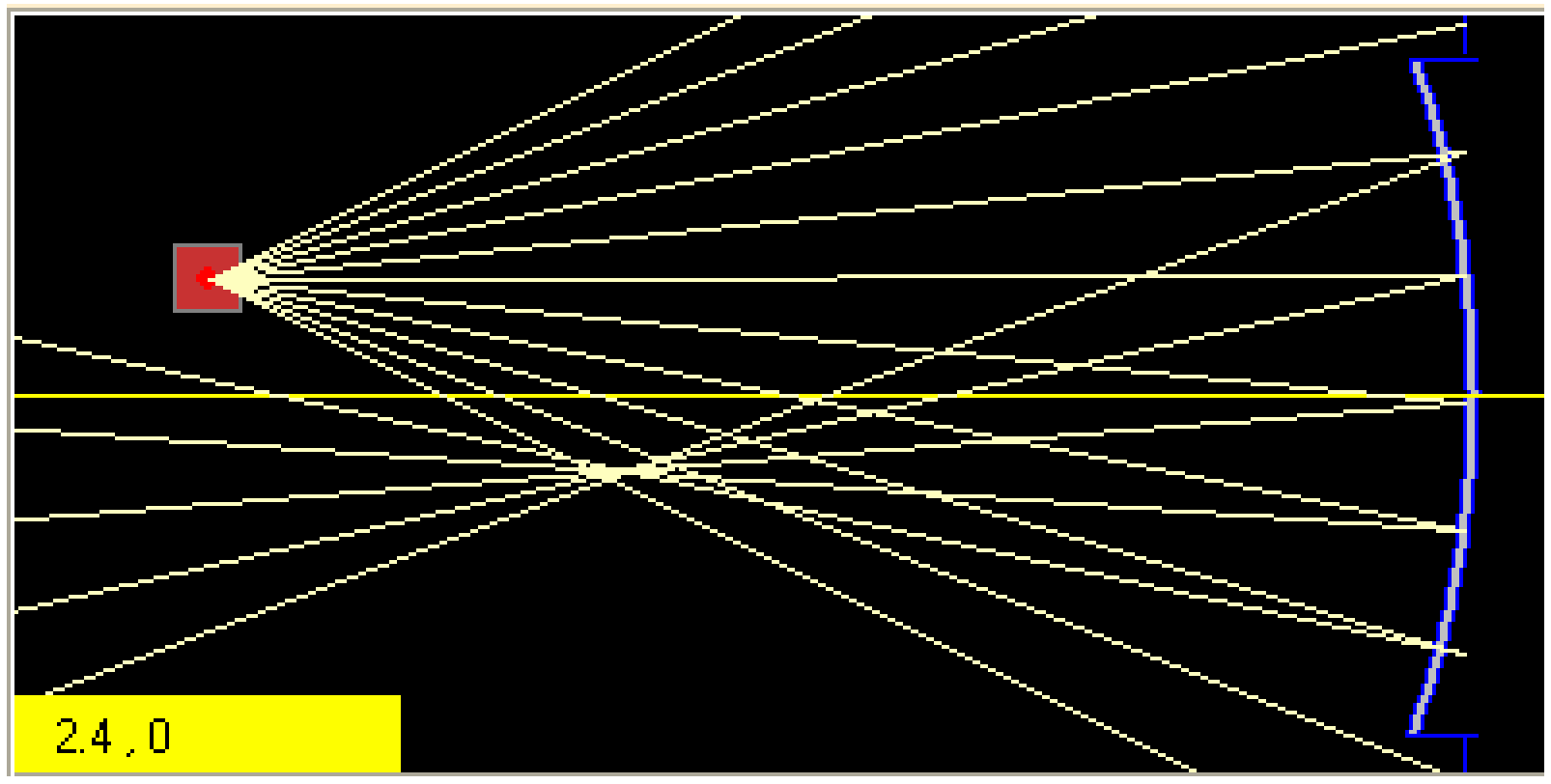

***Figure 1***: *Find the focal length of the mirror. Mouse coordinates are given in the yellow box.*

Notice that students are not given any variables. They must determine what information is relevant to answer the question (Where does an incoming parallel ray cross the principal axis?) and then interact with the Physlet to get that information (in this case, at x = 2.4 cm). They can then use this information to solve the problem. The student must also find the location of the mirror to determine the focal length. A "plug-n-chug" approach will not be effective in this problem as there is nothing to "plug" until some conceptualization of the problem has been done. It is also important to note that while the problem contains no overt "givens", it provides the means to access a plethora of information, most of which is unnecessary for the solution. Just as with real-world problems, the students must decide how to focus their efforts and what information to ignore.

Now, you may argue that the Physlet is unnecessary. After all, the problem can be asked and solved with just a picture and a ruler. But such an argument misses an important point. Students can not interact with a static picture. We have seen the benefit of interactivity in three very specific ways: the ability of the students to be creative, the learning opportunities provided, and the likelihood of students demonstrating misconceptions.

In the above example a solution was suggested based on finding where a parallel ray crossed the principal axis. If a static picture was all students had to work with, then the above solution would be the only method available. However, when working with the Physlet, students have the freedom to look at the problem from alternative angles. If a class is presented with this problem it is likely that a portion of the class will obtain a correct answer by moving the source along the principal axis to the point where all rays leaving the source are reflected parallel, thereby identifying the focal point. There are also other methods a student could use to obtain a correct answer. Physlet problems allow for creativity on the part of the student that is difficult to replicate in a paper-based problem.

We have also found Physlet-based problems to be a benefit to students because of the increased visualization. In the above example students did not need to move the source, but they could (and, in fact, virtually all of them will). As they interact with the applet they are seeing the effect of their actions. This allows for learning to take place that is somewhat separate from the solving of a specific problem.

The Physlet problem is often superior to a paper-based problem because the ability to interact with the applet gives students many different tracks to follow and hence improves assessment. For example, a portion of students encountering the problem discussed above will arrange the source in some version of what is shown in Figure 2.

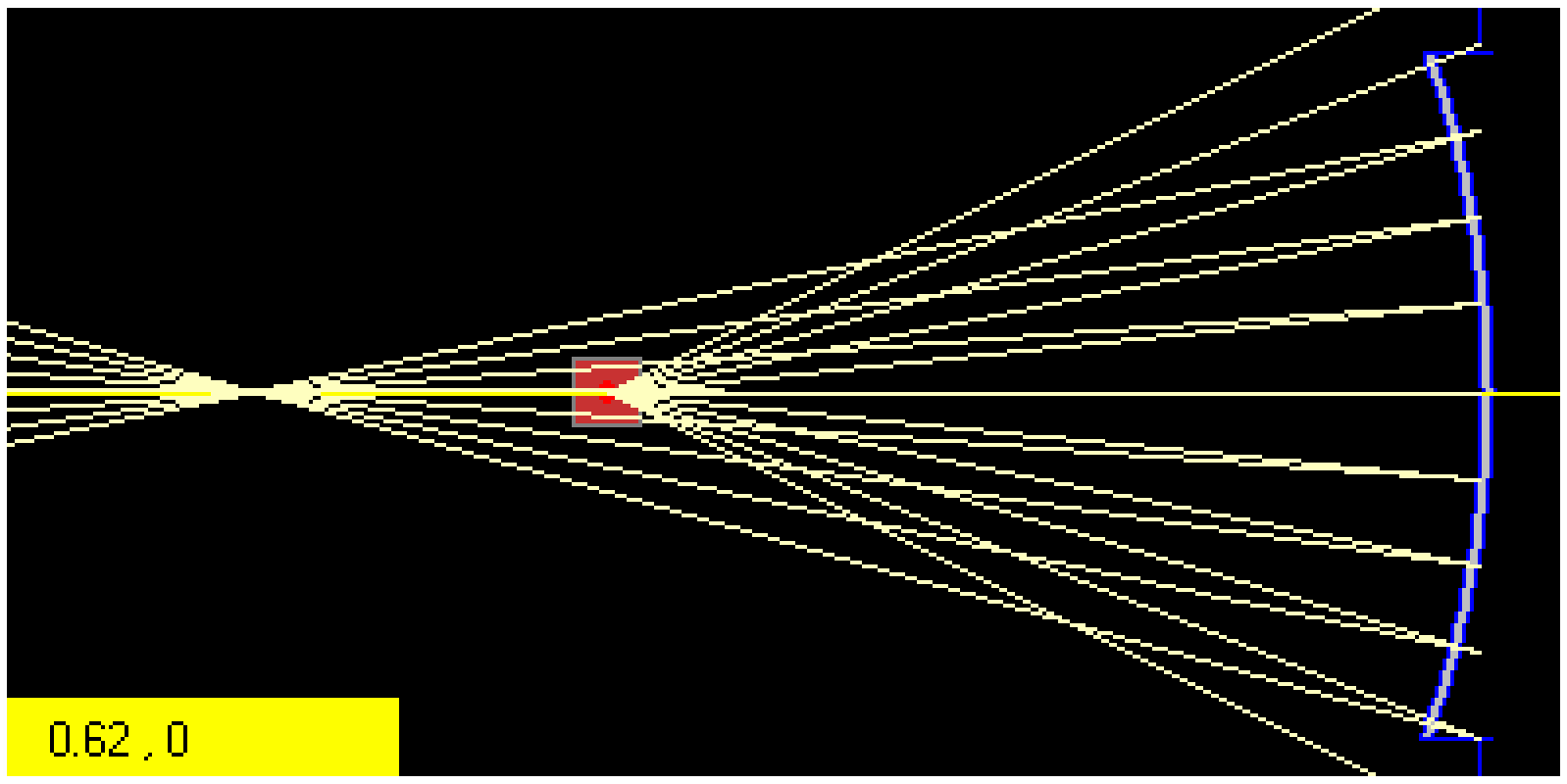

***Figure 2**: A point source on the principal axis with rays converging at x = 0.62 cm.*

The students will then declare that x = 0.62 cm is the focal point since the rays converge at this point. However, these students are incorrect because x = 0.62 cm is not where *parallel rays* would converge which is how a focal point is defined. Had the students been given a traditional problem that did not allow them to interact, this misconception would not have been so apparent. It should also be noted that students with this misunderstanding could easily dispel it themselves by noting that the point of convergence changed as the point source was moved. Since the location of the focal point is constant this should help them to refine their understanding.

We believe Physlet-based problems offer advantages over paper-based problems even when the Physlet is not integral to solving the problem. However, the dynamic and interactive nature of Physlet-based problems allows for the creation of problems that could not be asked using paper alone. Research[4,16] indicates that the greatest benefit of Physlet-based problems and questions will come when interacting with the Physlet is essential to the solution.

## Further Examples from Optics

*What is Behind the Curtain?*

We have developed a number of problems where students are given a source (point, infinite beam, or object) and a region hidden from view. They are asked to determine what optical element is behind the curtain. In the example shown in Figure 3, students are shown four such regions, asked to identify what is behind each curtain, and then asked to rank the objects in terms of their focal lengths.[17] Students can move and resize the sources.

In this particular problem, different types of sources are used so that students must utilize their understanding of each. For Curtain A, students must change the location of the object and note that the image is always on the same side of the mystery element as the object. At this point it is reasonable for the students to guess that a diverging lens is behind the screen. Further manipulation of the location and height of the object confirms this assertion. For Curtain B, students need to note that the element has no effect on the light. There is nothing behind Curtain B. Curtains C and D both contain converging lenses which students could deduce by observing the direction incoming rays are bent.

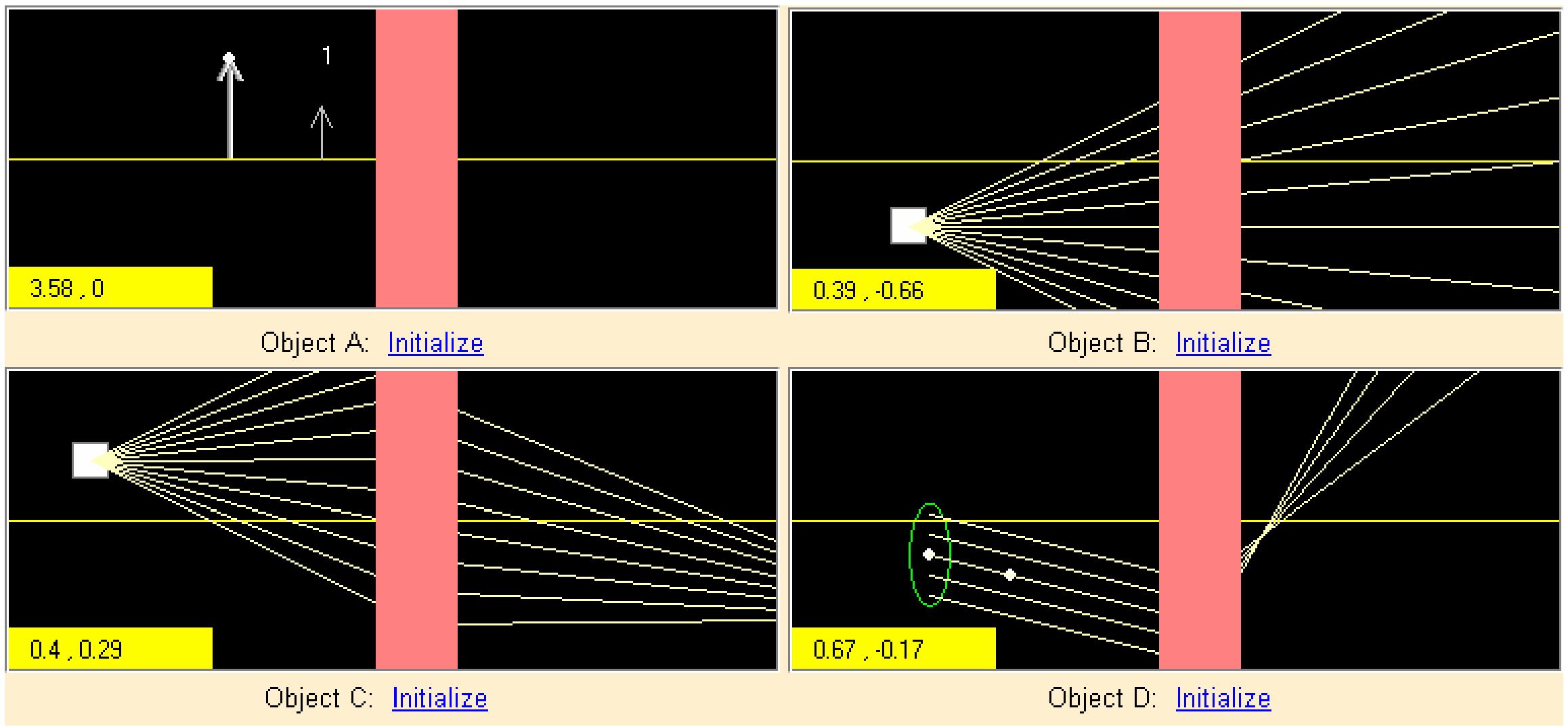

***Figure 3**: a) What is behind each curtain? b) Rank the objects in terms of their focal lengths from smallest to greatest.*

In order to rank the focal lengths of the mystery objects, students must first recognize that the focal length of the diverging lens is negative and that it is positive for the two converging lenses. This makes it unnecessary to calculate the actual focal length of the diverging lens, though it could easily be done by noting the locations of the object and image and using the lens equation. The focal lengths of the two converging lenses can be compared by noting where parallel rays converge.

This type of problem can be used to probe students' understanding of geometrical optics. It is easily modifiable, resulting in an almost limitless number of similar problems that could be created and tailored to a specific classroom.

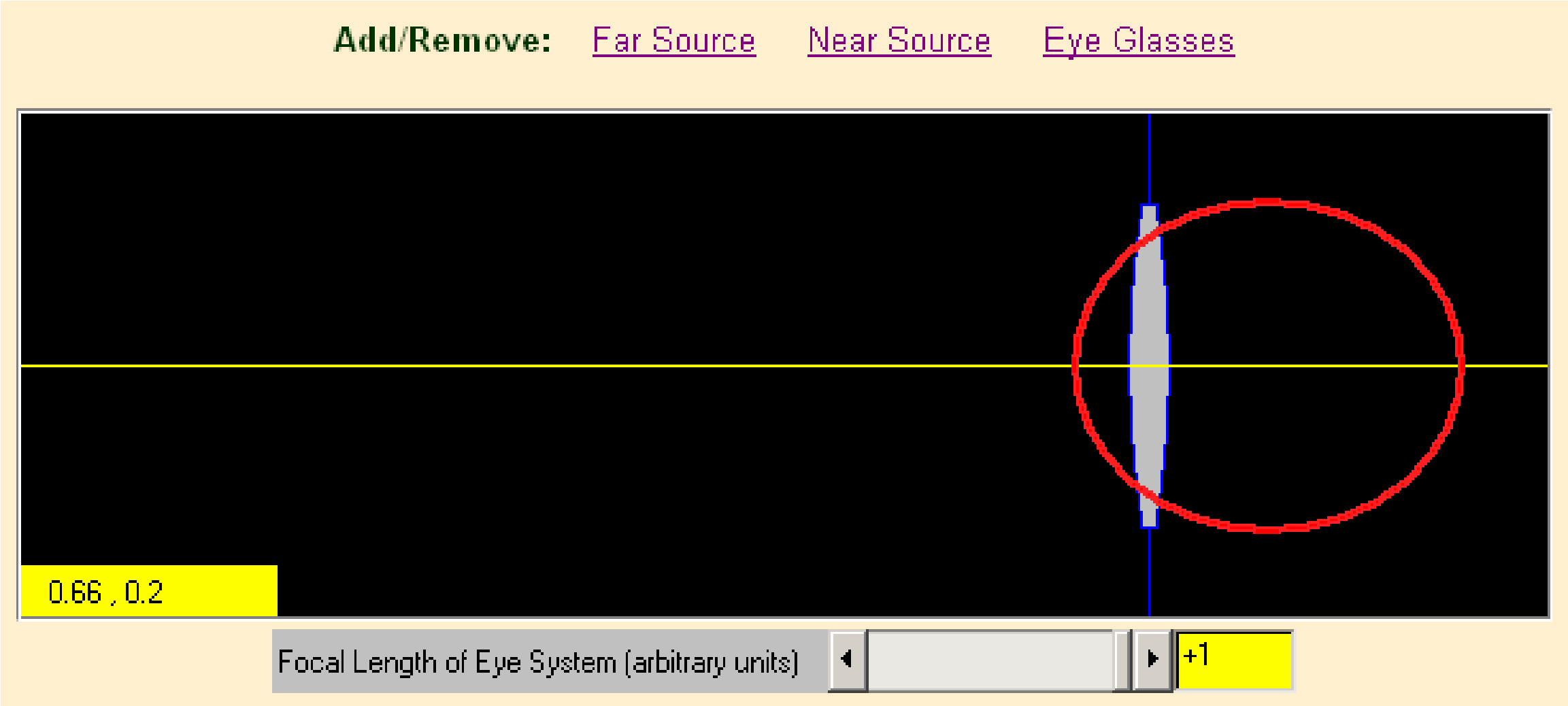

***Figure 4**: Is the eye represented in the above applet normal, nearsighted or farsighted? What power eyeglasses should be used for normal vision (can see far away objects with eyes relaxed, and also focus on an object at x=2.0, in arbitrary units).*

*A Model of the Eye*

Students are usually motivated by problems they see as relevant to their everyday lives. Our students found the Physlet problem shown in Figure 4, relating to the eye and corrective lenses, to be exciting and relevant, and claimed it helped them to visualize the functioning of the eye. In this question, students are asked to check vision and prescribe eyeglasses. Students can add near or far sources of light and see the interaction with the eye. They can use the slider at the bottom of the applet to change the focal length of the lens system of the eye to simulate the eye's ability to accommodate. The eye is not shown to scale and units are arbitrary because the actual scale is too small to allow the functioning of the eye to be noticeable.

In order to answer the question, students must determine if a far and/or near object can be focused on the back of the retina. (Figures 5a, 5b, and 5c) In this case the person can focus on near objects but not far objects so he/she is nearsighted. Then, a lens simulating eyeglasses must be inserted to determine the type of lens and its focal length. (Figure 5d) Students can then verify that near objects can still be seen with the eyeglasses on.

In order to answer this question correctly, students must have a strong conceptual understanding of the operation of lenses, especially as that operation relates to the optical system of the eye. Students who have this understanding can easily answer this question. Students who have misunderstandings will find it difficult to answer the question. We used this question on an exam and found it to be beneficial for assessing student understanding.

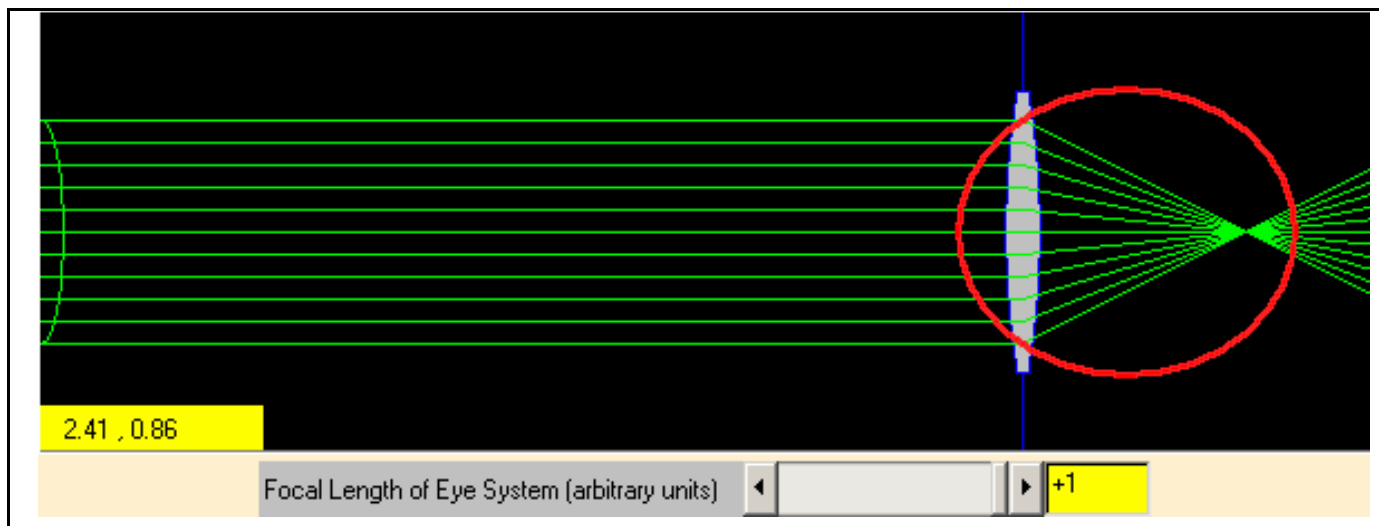

***Figure 5a**: Light from very far away is focused in front of the retina.*

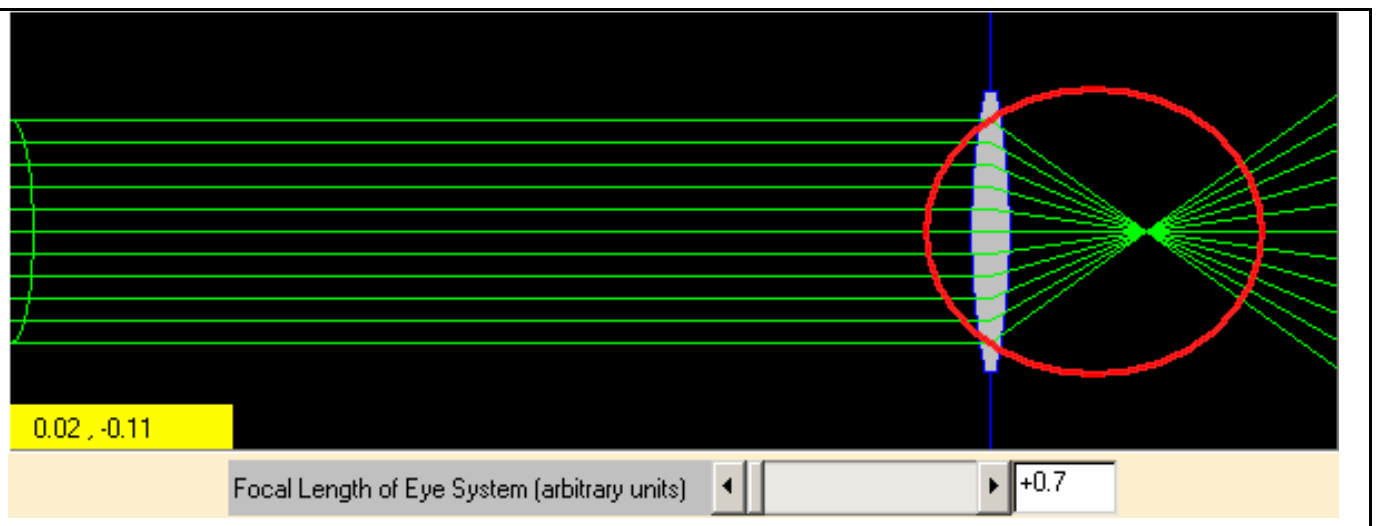

***Figure 5b**: The eye can not accommodate to focus the light at the back of the retina.*

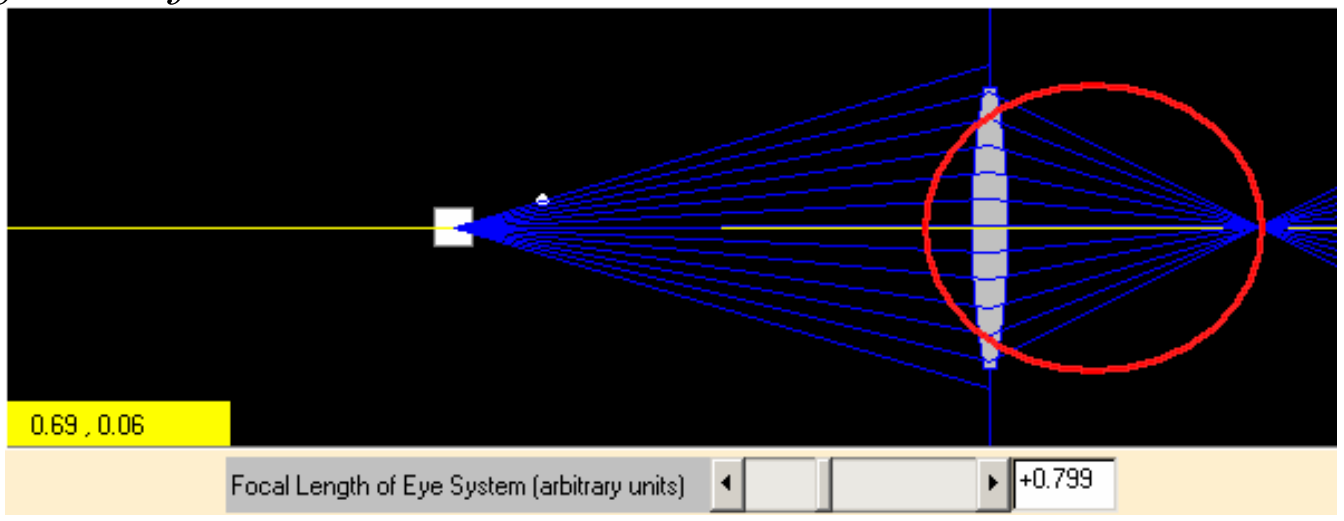

***Figure 5c**: A source at the near point can be focused at the back of the retina.*

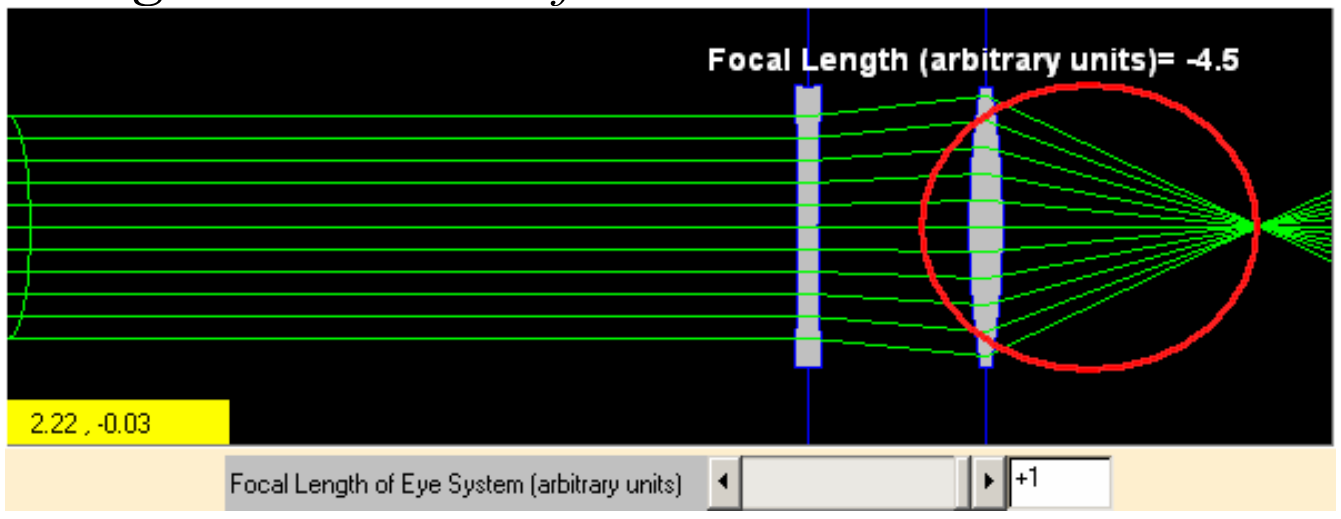

***Figure 5d**: Eyeglasses with a diverging lens of focal length= -4.5 (arbitrary units) can be used to focus light from a far away source at the back of retina.*

*Index of Refraction*

Another type of question that can be asked involves the index of refraction. In this particular example, shown in Figure 6, students are presented with an applet that displays a point source and a lens. The index of refraction of the medium surrounding the lens can be changed. As the index of refraction is changed, the color of the background varies as a visual clue to the index change. Students are then asked to find the index of refraction of the lens. In order to determine the index of refraction of the lens, students can change the index of refraction of the medium surrounding the lens until the light passing through the lens is unaffected. At that point, the index of refraction of the lens must be the same as the index of refraction of the surrounding medium.

Although this is a fairly easy problem for students, the process of solving the problem can help students to develop their understandings of refraction and lens operation. The problem reinforces the idea that lenses work by means of a different index of refraction from the surrounding medium. It can also be used to help students see that a "converging" lens is only converging if placed in a medium where the index of refraction of the medium is less than that of the lens. A glass lens in air will be converging, but if the same glass lens is placed in a medium of higher index of refraction it will act as a diverging lens. These issues are important for developing a conceptual understanding of lenses but are often glossed over in traditional instruction.

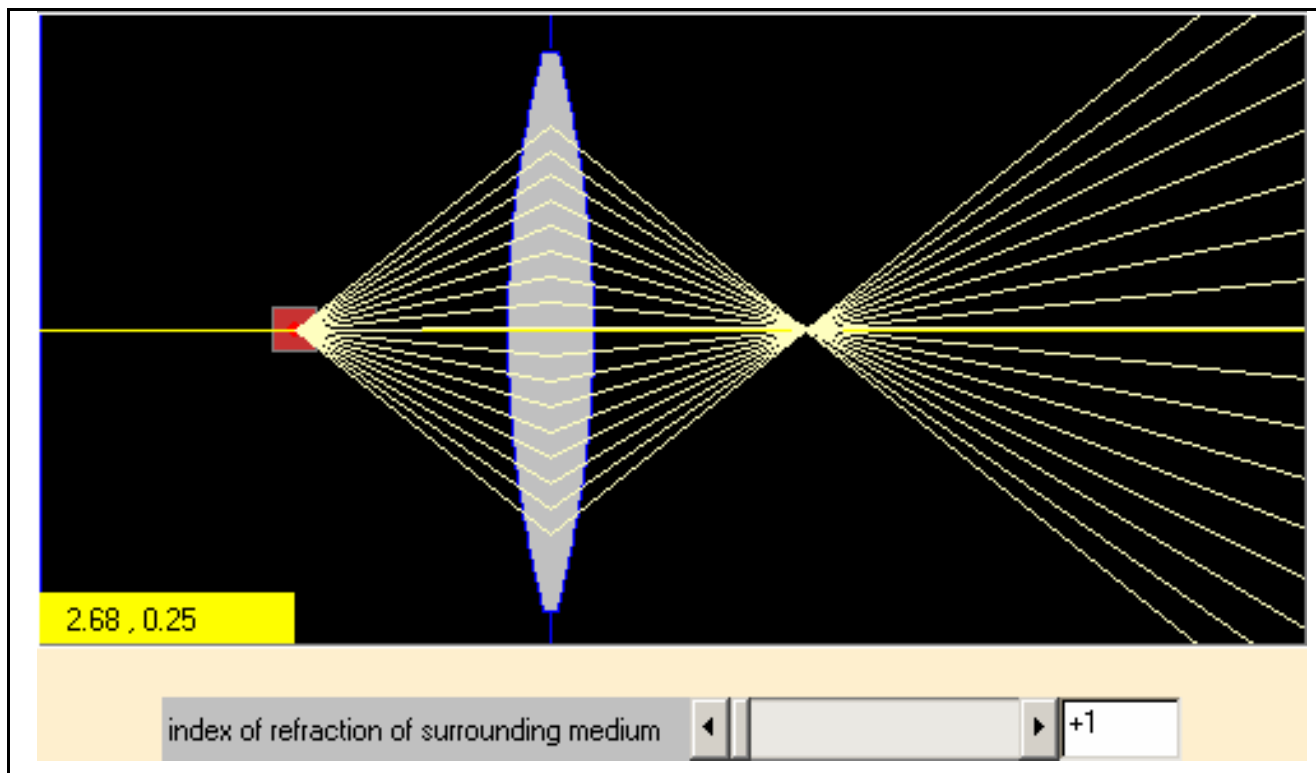

***Figure 6a***: *What is the index of refraction of the lens?*

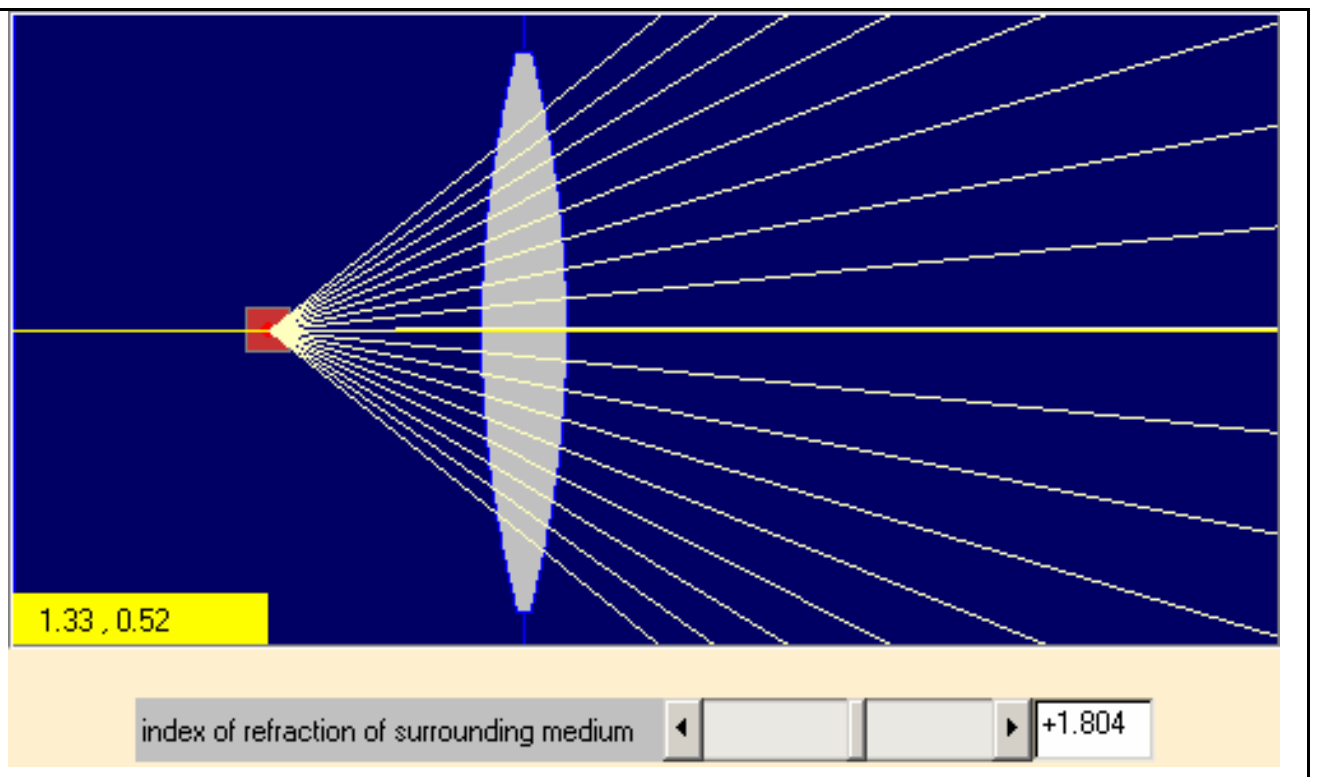

***Figure 6b***: *When the index of refraction of the surrounding medium is 1.8, light is unaffected as it passes through the lens.*

*Ripple Tank Simulation*

The last example of an optics-based Physlet is not a problem but rather a simulation that can be used as a traditional ripple tank. Unlike a physical ripple tank, the applet creates no mess and can be made accessible to students outside of class. The interface of this demonstration is shown in Figure 7.

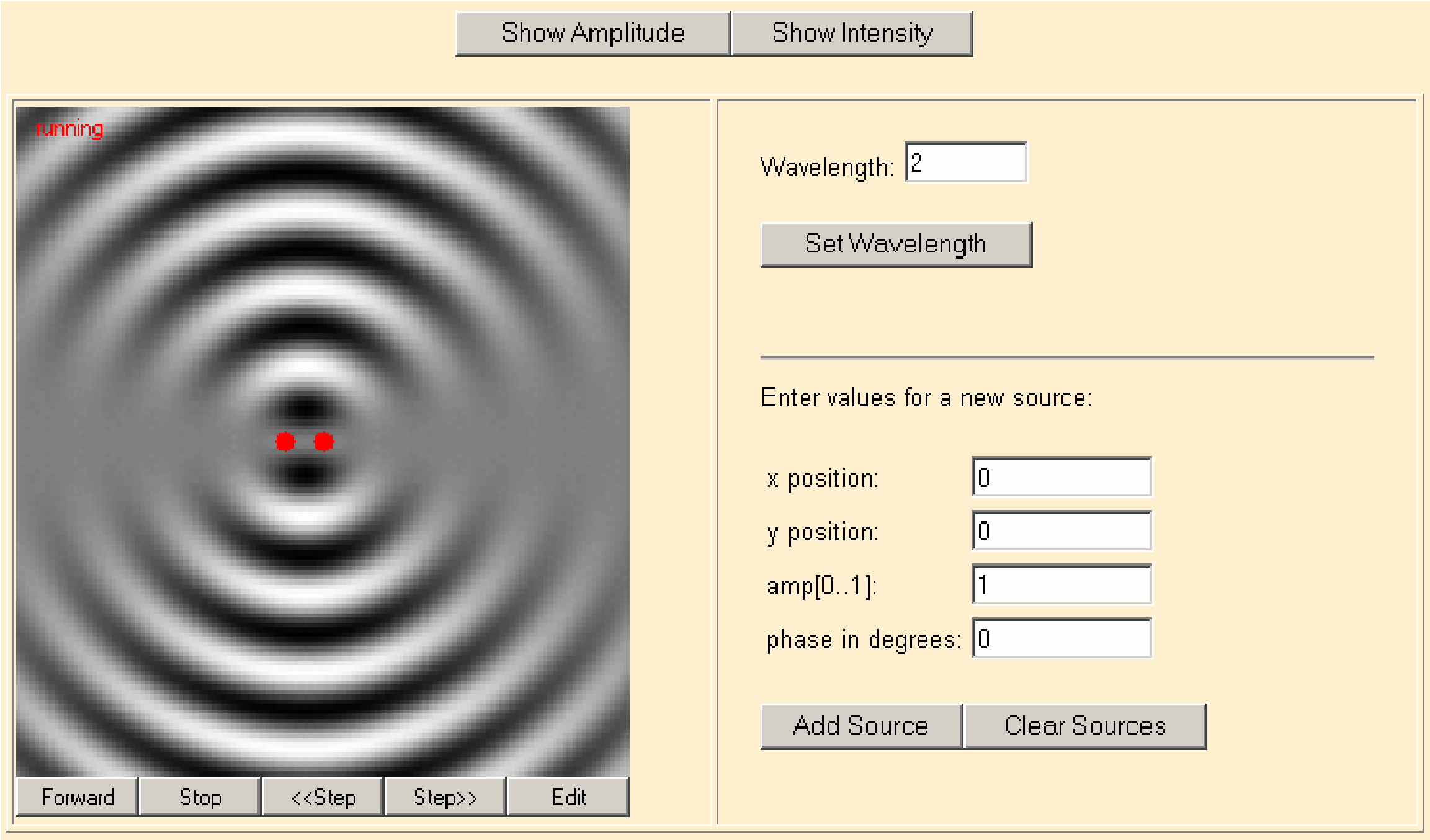

***Figure 7***: *An applet to demonstrate interference effects.*

The demonstration can be used in class or as an out of class tool for students to play with as they construct their understandings. It can also be used to construct a number of related problems.

The applet allows the user to create any number of wave sources, put them at any location, and specify their amplitude and phase. The wavelengths of all the sources must be the same, but can be set to a desired value. When the applet plays, waves can be seen moving away from the sources, just as with a ripple tank.

A number of questions can be derived from this applet regarding interference effects. For example, we use the ripple tank demonstration to create an interference pattern and then ask the students to determine the relative phase of the sources based on the pattern generated. Alternatively, students could be asked to determine the separation of sources in terms of the wavelength of light by considering the pattern generated. Students often have tremendous difficulty connecting interference effects with a difference in path length.[18] The ripple tank can be used to give them practice in analyzing interference effects separate from the double slit experiment, whose results they often memorize and apply without understanding.

**Obtaining the Problems**

All of the problems mentioned in this article, as well as many other resources, are available on the Physlet resources site.[19] The examples in this paper can be found in the link to "Davidson Optics" under collections. In addition, many other Physlet problems and questions are found on the general Physlet website and on the CD that accompanies the previously mentioned *Physlets* book. Once the problems are obtained they must be properly set up to run. Instructions for how to do this can be found on the Physlet website and/or in the Physlets book. Although the Physlet name is a registered trademark, Physlets and Physlet problems can be used without charge for non-commercial use.

We are currently developing an extensive collection of ready-to-run Physlet-based problems and questions covering all topics in introductory physics. We expect this collection, which will include the problems discussed above, to be available in the "*Physlet Workbook*" from Prentice Hall by August 2003.

**Acknowledgements**

The Physlet project has been generously supported by NSF (DUE-9752365 and DUE-0126439). We could also like to thank Ken Krebs for allowing us to field-test some of the problems, and Larry Cain and Aaron Titus for their helpful feedback. Also, the Davidson College physics department has provided much support and encouragement to the Physlet project.

---

[1] M. T. H. Chi, P. S. Feltovich, and R. Glaser, "Categorization and Representation of Physics Problems by Experts and Novices," *Cognitive Science*. **5,** 121-152 (1981).
[2] J. Larkin, J. McDermott, D. P. Simon, and H. A. Simon, "Expert and Novice Performance in Solving Physics Problems, *Science*, **208**, 1335-1342 (1980).
[3] A. Van Heuvelen, "Learning to Think Like a Physicist: A Review of Research-Based Instructional Strategies," *American Journal of Physics*, **59**, 891-897 (1991).
[4] A. Titus, "Integrating Video and Animation with Physics Problem Solving Exercises on the World Wide Web." Doctoral Dissertation, 1998.

[5] G. Polya, *How to Solve It,* 2nd ed. (Doubleday, Garden City, NY, 1957).
[6] W. Leonard, R. Dufresne, and J. Mestre, "Using Qualitative Problem-Solving Strategies to Highlight the Role of Conceptual Knowledge in Solving Problems," *American Journal of Physics,* **64**, 1495-1503 (1996).
[7] R. Serway and R. Beichner, *Physics for Scientists and Engineers, 5th ed.* (Saunders, Philadelphia, 2000), p. 47.
[8] F. Reif, J. H. Larkin, and G. C. Brackett, "Teaching General Learning and Problem-Solving Skills," *American Journal of Physics*, **44**, 212-217 (1976).
[9] A. Van Heuvelen, "Overview, Case Study Physics," *American Journal of Physics*, **59**, 898-907 (1991).
[10] P. Heller, R. Keith, and S. Anderson, "Teaching Problem Solving Through Cooperative Grouping. Part 1: Group Versus Individual Problem Solving." *Journal of Physics*, **60**, 627-636 (1992).
[11] Estimation problems are also known as Fermi problems. A large collection is maintained by the PER group at the University of Maryland. http://www.physics.umd.edu/rgroups/ripe/perg/fermi.html
[12] S. W. Bonham, J. S. Risley, and W. Christian, "Using Physlets to Teach Electrostatics," *The Physics Teacher*, **37**, 276-280 (1999).
[13] M. Belloni, L. Cain, and W. Christian, "*Using Just-In-Time Teaching and Physlets in Quantum Mechanics,"* talk given at American Association of Physics Teachers, Rochester NY, 2001.
[14] W. Christian, and M. Belloni, *Physlets: Teaching Physics with Interactive Curricular Material,* (Prentice Hall, Upper Saddle River, NJ, 2001).
[15] http://webphysics.davidson.edu/Applets/Applets.html Maintained by the Davidson College physics department.
[16] M. Dancy, "Investigating Animations for Assessment with an Animated Version of the Force Concept Inventory." Doctoral Dissertation, 2000.
[17] T. L. O'Kuma, D. P. Maloney, and C. J. Hieggelke, *Ranking Task Exercises in Physics*, (Prentice Hall, Upper Saddle River, NJ, 2000).
[18] K. Wosilait, P. Heron, P. Shaffer, and L. McDermott, "Addressing Student Difficulties in Applying a Wave Model to the Interference and Diffraction of Light," *American Journal of Physics Supplement*, **67**, S5-S15, (1999).
[19] http://webphysics.davidson.edu/physlet_resources/